## Title

LitUbik: open-source software for human-led systematic reviews with traceable AI assistance

## Authors

Adolphe J. Béquet[a,*], Sébastien Laurent[b], Baptiste Bouvier[a]

[a] Laboratory of Ergonomics and Cognitive Science for Transportation (LESCOT), Gustave Eiffel University, France

[b] Independent web developer, Lyon, France

* Corresponding author. E-mail: adolphe.bequet@univ-eiffel.fr

## Abstract

LitUbik is an open-source, self-hosted tool for conducting systematic reviews, from the registered protocol to the PRISMA 2020 diagram. LitUbik covers calibrated search equations, automated searching, deduplication, blinded multi-reviewer screening, full-text retrieval, snowballing, appraisal, extraction and reporting, on one server or across disconnected instances. At each stage, optional AI assistance from commercial or local language models offers proposals that a named reviewer accepts or rejects. Every AI call is logged, so the AI-use disclosure is written from a record. A systematic review of 31 papers on review tools, conducted in LitUbik, compares it with existing tools and illustrates its use. LitUbik is distributed under AGPLv3.

## Metadata

| Nr. | Code metadata description | Metadata |
|---|---|---|
| C1 | Current code version | v1.0.0 |
| C2 | Permanent link to code/repository used for this code version | https://github.com/abequet/litubik_public, archived on Zenodo: https://doi.org/10.5281/zenodo.23652584 |
| C3 | Legal code license | GNU Affero General Public License v3 or later (AGPL-3.0-or-later) |
| C4 | Code versioning system used | Git |
| C5 | Software code languages, tools and services used | Python 3.12+, FastAPI, SQLAlchemy 2 and Alembic, SQLite (WAL), React 19, TypeScript, Vite, Tailwind CSS, pdf.js, PyMuPDF, rapidfuzz, scikit-learn, axe-core, Node.js 24 (frontend build), Playwright (optional, browser-based and institutional retrieval). Optional LLM APIs (Anthropic, OpenAI, Google, xAI, Mistral) or any OpenAI-compatible local server. Docker (optional). |
| C6 | Compilation requirements, operating environments and dependencies | Windows, macOS, Linux. On Windows: packaged executable (no prerequisites) or one-click source launcher (Python ≥ 3.12, Node.js downloaded automatically). On any system: Docker Compose, or installation from source (Python ≥ 3.12, Node.js 24). Online demonstration, without installation or account: https://demo.litubik.org |

| Nr. | Code metadata description | Metadata |
|---|---|---|
| C7 | If available, link to developer documentation/manual | https://github.com/abequet/litubik_public/tree/v1.0.0 (README.md, and docs/help/ in English and French) |
| C8 | Support email for questions | adolphe.bequet@univ-eiffel.fr |

## 1. Motivation and significance

Systematic literature reviews (SLRs) are a cornerstone of evidence synthesis, and methodological expectations keep rising. Reviews are expected to report according to PRISMA 2020 [1] and its search extension PRISMA-S [2]. Question frameworks such as PICO, PICOS and SPIDER organise the search around the concepts of the review question [3]. PRISMA 2020 asks reviews to report their registration [1], and platforms such as OSF provide time-stamped registrations [4]. Cochrane requires two independent reviewers for inclusion decisions [5]. Software engineering guidelines advise reporting Cohen's κ when several reviewers screen [6,7]. Ethics bodies and publishers now ask authors to disclose how they used AI [8,9]. Each stage produces artefacts: search strings and hit counts, deduplication logs, reviewer decisions, conflict resolutions, exclusion reasons, appraisal scores and AI calls. These must stay consistent for months, because the numbers in the manuscript come from them.

In practice, teams stitch the process together from separate tools. They search in database interfaces, deduplicate in reference managers, record decisions in spreadsheets and draw the PRISMA diagram by hand. No single tool covers the requirements that reviewers rank highest [10]. Reporting suffers: few software engineering reviews report their search dates, and several mix counts of papers and studies [11]. A single changed decision can silently invalidate the PRISMA counts of a manuscript [12].

Three further tensions have sharpened. First, commercial platforms charge subscriptions. In 2022, Covidence cost US$240 per review per year and DistillerSR US$75 per reviewer per month [13]. Second, restrictive data policies or embargoed submissions may forbid uploading records to a cloud service, which motivates locally deployable tools [12]. Third, large language models (LLMs) have entered evidence synthesis. Early evaluations found screening performance that varied widely across reviews, prompts and decision rules [14–16]. A scoping review found that validated LLM applications for conducting reviews are still lacking [17]. LLM output is also hard to reproduce: two nominally identical screening runs disagreed on 94 of 1,131 records, 29 of them eligible [18]. These results favour LLMs as supervised assistants rather than autonomous reviewers [18]. Cochrane, Campbell, JBI and the Collaboration for Environmental Evidence now expect human oversight of AI, and full reporting of any AI that makes or suggests judgements [19]. Self-reports are a weak basis for that reporting, because people misremember what they produced with AI [20].

To position LitUbik, we used it to conduct a systematic review of review-support tools (Section 3). Thirty-one papers published between 2013 and 2026 describe these tools [12,13,21–49]. Our feature matrix compares 14 multi-stage workbenches, 14 single-stage or generative tools and a family of commercial discovery assistants. Table 1 condenses it for the ten most comparable workbenches. An earlier feature analysis already found weak search and collaboration support in such tools [50]. Five findings structure this paper.

First, most tools start after the search. LitFlow searches several databases through their APIs [26]. Thoth searched Scopus only in its first version [25], and DistillerSR and EPPI-Reviewer query PubMed only [27]. None calibrates an equation against a set of must-find references.

Second, the stages after screening are rarely automated. Among the ten tools, only DistillerSR retrieves full texts automatically [27]. A review of 21 AI-enhanced tools found none with automated snowballing and four that draw the PRISMA flow diagram [27].

Third, blinded dual screening and live agreement statistics rarely coexist. Covidence enforces blinded screening [29] and reports agreement for screening only [13]. ReviQ computes κ and PABAK, but its reviewers see each other's names [12]. None of the ten reports Fleiss' κ for larger teams.

Fourth, AI assistance is absent, opaque or unlogged. ReviQ and Thoth 2.0 list it as future work [12,24]. Ranking models give each record a relevance score without criterion-level explanation [21,23,27]. Only LitFlow gives criterion-level LLM suggestions with confidence and rationale [26]. No tool logs its own AI calls for disclosure, the log-based approach that the DAISY framework describes but leaves to future study [51].

Fifth, collaboration is either cloud-mediated or a manual file exchange. ReviQ's author reports that its JSON exchange works for two or three reviewers but does not scale to larger groups [12].

Table 1 also shows what LitUbik lacks. LitUbik offers every feature that the sources report for at least one of the ten tools, with five exceptions. It has no hosted production service (demonstration only: https://demo.litubik.org), no mobile app, no machine-learning prioritisation, no RCT classifier and no published user evaluation. None of the ten tools documents accessibility conformance, which library guidance asks institutions to check before adoption [32].

LitUbik was developed to close these gaps in one self-hosted application. Its design rests on one principle: the reviewers conduct the review, and AI assists them. At each stage, an optional LLM can propose a structured protocol, search terms, screening assessments, extraction values or report drafts. Each proposal waits for a named reviewer to accept, edit or reject it, and every billed call is logged. LitUbik supports the PICO, PICOS, SPIDER, SPICE and PEO frameworks, or a free format. It connects to commercial LLM providers or to a local model run by the team, but ships no model of its own. It enforces double blinding on the server and computes Cohen's and Fleiss' κ [7,52]. Sections 2 to 4 describe the software, the review conducted with it, and its impact and limitations.

**Table 1.** Features of ten review workbenches, as reported in their published descriptions and read during the review of Section 3. Sources: Covidence [13,27,29,31,33], DistillerSR [13,27,28,31,33], EPPI-Reviewer [13,27,31], Rayyan [21,27,29], CADIMA [28,31], StArt [12,22,31], LitFlow [26], Thoth 2.0 [24,25], ReviQ [12], ASReview [23,27]. ✓ = reported, (✓) = partial or with reservations, blank = not reported in the sources. A cell describes a tool as its sources describe it, and tools may have evolved since. The 2018 sources predate PRISMA 2020. Recall estimation is offered by SWIFT-Active Screener [30], outside the table. LitUbik column: code as of October 2026.

| Feature | Covidence | DistillerSR | EPPI-Reviewer | Rayyan | CADIMA | StArt | LitFlow | Thoth 2.0 | ReviQ | ASReview | LitUbik |
|---|---|---|---|---|---|---|---|---|---|---|---|
| **Deployment and licence** | | | | | | | | | | | |
| Open source | | | | | | | | ✓ | ✓ GPLv3 | ✓ Apache | ✓ AGPLv3 |
| Hosted service, no installation | ✓ | ✓ | ✓ | ✓ | ✓ | | (✓) | ✓ | | | (✓) demo |
| Runs locally or self-hosted | | | | | | ✓ desktop | (✓) container | (✓) Docker | ✓ Docker | ✓ | ✓ exe, server, Docker |
| Mobile app | | | | ✓ | | | | | | | |
| **Protocol and search** | | | | | | | | | | | |
| Structured protocol with coded criteria | (✓) | | | | ✓ | ✓ | (✓) | ✓ | ✓ | | ✓ five frameworks |
| Import of a registered protocol | | | | | | | | | | | ✓ |
| Search-equation builder | | | | | | (✓) | (✓) | ✓ | | | ✓ |
| Translation to each database's syntax | | | | | | | (✓) | (✓) | | | ✓ 19 databases |
| Automated searching via APIs | | (✓) PubMed | (✓) PubMed | | | | ✓ | (✓) Scopus | | | ✓ 8 open, 6 keyed |
| Equation calibration on must-find references | | | | | | | | (✓) string metrics | (✓) precision, recall | | ✓ recall per database |
| Deduplication | (✓) | ✓ | (✓) | (✓) | (✓) | (✓) | ✓ | (✓) | ✓ DOI, title | | ✓ fuzzy, arbitration |
| Living review (scheduled update passes) | | | | | | | | | | | ✓ |
| **Screening** | | | | | | | | | | | |
| Dual screening with enforced blinding | ✓ | (✓) | (✓) | (✓) optional | (✓) | | | (✓) | (✓) names visible | | ✓ server-side |
| Live inter-rater agreement | (✓) | | | | (✓) pilot κ | | (✓) AI vs human | | ✓ Cohen, PABAK | | ✓ Cohen, Fleiss |
| ML prioritisation or active learning | (✓) | ✓ | ✓ | ✓ | | (✓) keyword score | | | | ✓ | |
| RCT or risk-of-bias classifiers | ✓ | | ✓ | | | | | | | | |
| LLM screening suggestions, human-validated | | | | | | | ✓ | | | | ✓ three levels |
| **Full text and snowballing** | | | | | | | | | | | |
| Automated full-text retrieval | | ✓ | | | | | (✓) open access | | | | ✓ OA, institutional |
| PDF viewer with annotations | | | (✓) | | | | (✓) | | | | ✓ shared |
| Automated backward and forward snowballing | | | | | | (✓) backward | | ✓ Crossref | (✓) manual | | ✓ 4 sources, network |
| **Appraisal, extraction and reporting** | | | | | | | | | | | |
| Quality appraisal | ✓ | (✓) | ✓ | | ✓ | ✓ | | ✓ | ✓ | | ✓ JBI, MMAT, own |
| Typed extraction / double extraction | ✓ / ✓ | ✓ / | ✓ / ✓ | | ✓ / (✓) | ✓ / | ✓ / | ✓ / | ✓ / | | ✓ / ✓ |
| PRISMA flow diagram from live data | ✓ | ✓ | | (✓) | ✓ | | | | ✓ | | ✓ per round |
| Methods and AI disclosure from audit trail | | | | | (✓) | | (✓) | (✓) | (✓) | | ✓ |
| Replication package or project archive | | | | | | (✓) | | | ✓ | | ✓ |
| Sync between disconnected instances | | | | | | | | | (✓) JSON | | ✓ |
| Accessibility conformance statement | | | | | | | | | | | (✓) WCAG 2.1 AA, partial |
| Published user evaluation | ✓ | (✓) | ✓ | ✓ | (✓) | (✓) | ✓ | ✓ | | ✓ | |

## 2. Software description

### 2.1. Software architecture

LitUbik is a client–server web application, designed so that one researcher's machine can serve a team. Its FastAPI (Python) backend stores data in SQLite (WAL mode, the only tested database) through SQLAlchemy 2 and Alembic. The backend also serves the React 19/TypeScript frontend. One port on a local network, or a Docker container behind a reverse proxy, therefore serves all collaborators. A packaged Windows executable, a one-click source launcher and a Docker Compose file cover three deployment scales (Fig. 1).

Four cross-cutting subsystems shape the design. First, users hold per-project admin or reviewer roles. Every methodological rule, such as vote blinding, arbitration rights or the write-locks of satellite instances, is enforced in the API layer rather than only hidden in the interface. Second, an in-process job runner executes long operations such as searches, AI batches, retrieval and snowballing, and streams their progress to the browser. Third, an LLM layer serves five commercial providers (Anthropic, OpenAI, Google Gemini, xAI and Mistral) and any OpenAI-compatible local server, such as Ollama or vLLM. API keys are stored encrypted. The project chooses the model for each of eleven tasks, to match model cost to task difficulty. Automatic failover between providers can be disabled for strict reproducibility. A cost estimate precedes each batch, and prompts are readable and editable per project. The AI-use journal tallies billed calls, tokens and cost per task, model, user and day. Fourth, the interface is bilingual (English and French). Its accessibility statement targets WCAG 2.1 AA, with two criteria partially supported. An automated axe-core audit of eleven main pages in both languages runs in continuous integration with 662 backend tests.

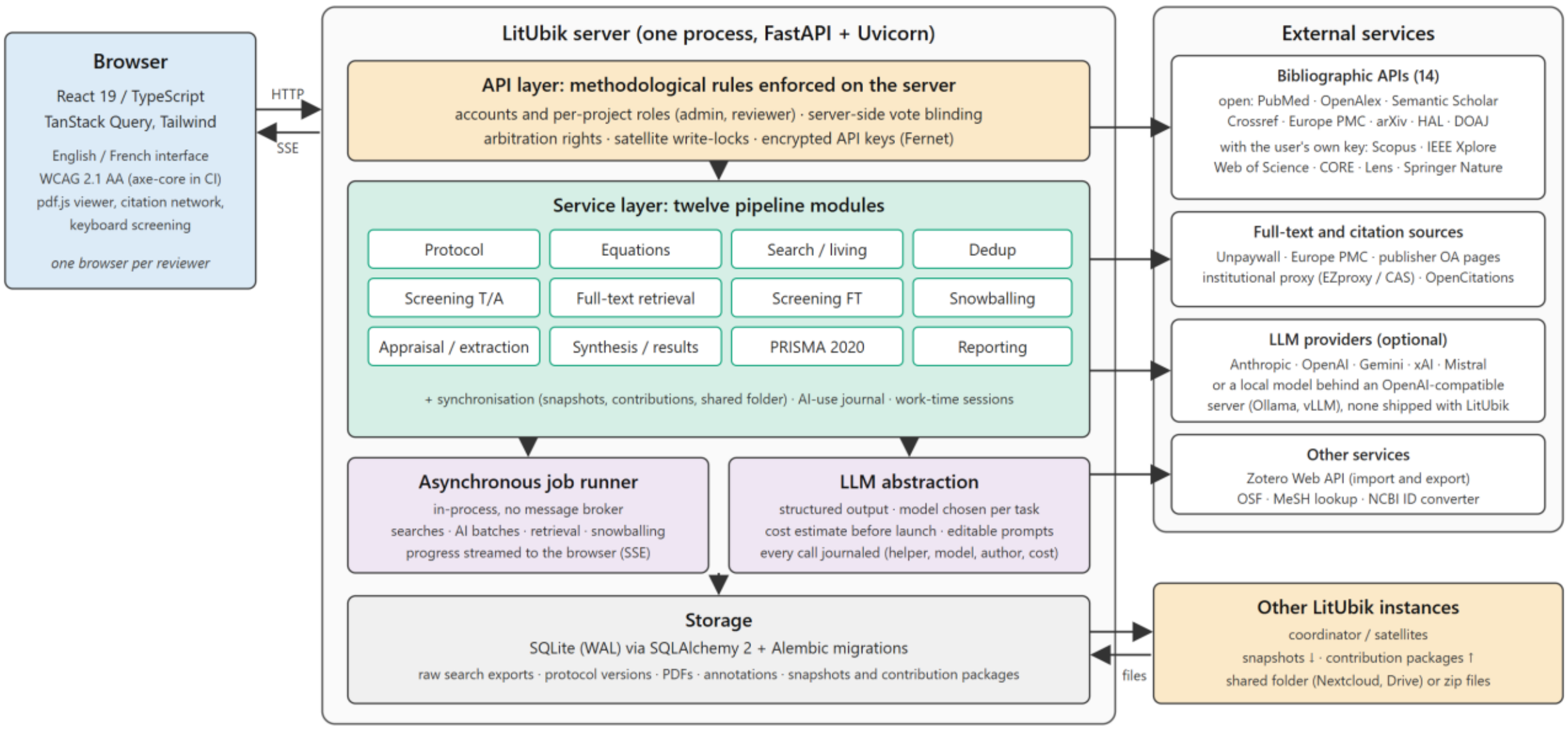

**Figure 1.** Architecture of LitUbik. One FastAPI process serves the React frontend, enforces the methodological rules in its API layer, runs the twelve pipeline modules and the asynchronous jobs, and calls the external services. Deployment: a packaged Windows executable, a one-click source launcher, or Docker Compose behind a reverse proxy. One port on a local network serves the whole team, and only calls to the configured external services leave the machine.

### 2.2. Software functionalities

LitUbik organises a review as twelve stages mapped onto the PRISMA 2020 phases (Fig. 2).

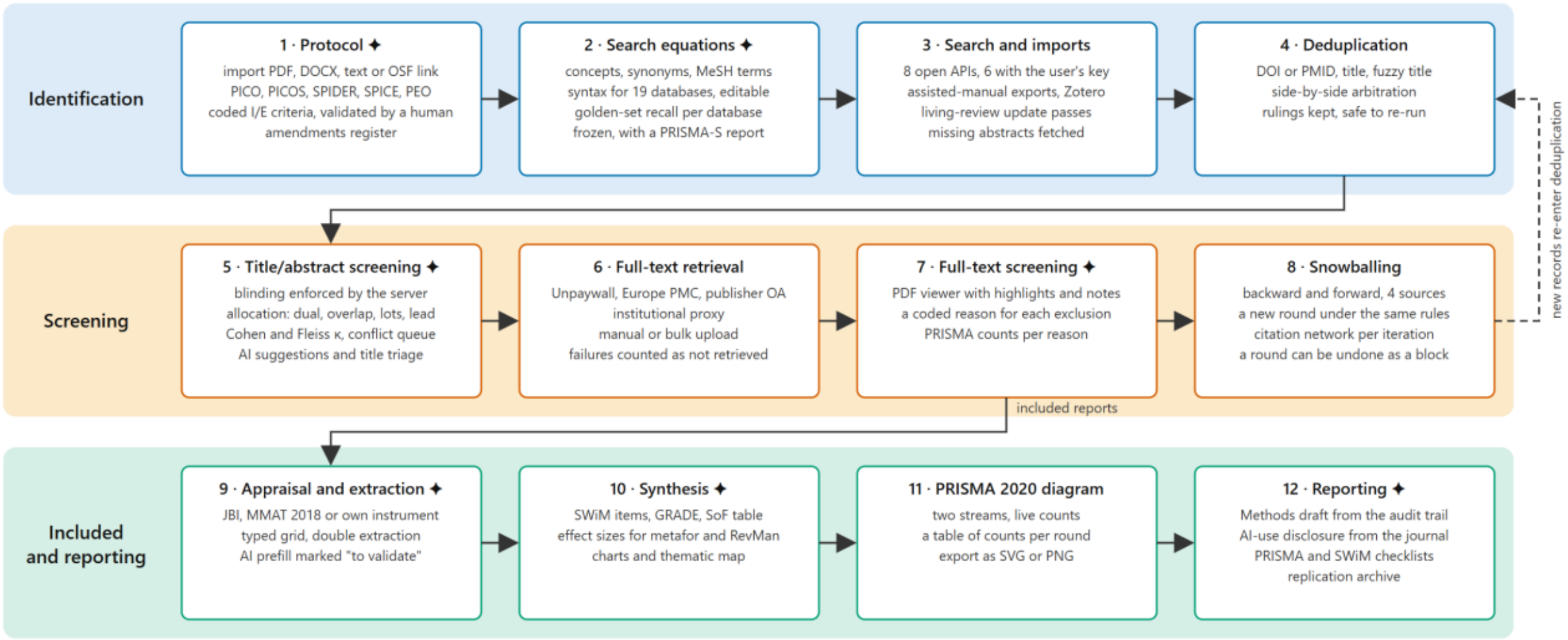

**Figure 2.** The twelve stages of a LitUbik review, read from left to right in the three PRISMA 2020 phases. ✦ marks the stages where an LLM can make proposals. A named reviewer validates each proposal, and the call is logged with its model, author and cost. Snowballing and living-review rounds re-enter at deduplication and repeat stages 5 to 7.

**Protocol.** A registered protocol is imported as PDF, DOCX, Markdown or text, or from an OSF link. Its registry and registration number are recorded for PRISMA 2020 item 24a [1]. An LLM proposes a structured version: background, questions in the chosen framework, and inclusion and exclusion criteria. The server gives each criterion a stable code (I1, E1, …). The methodologist edits this proposal and must validate it. The validated protocol then frames every later AI prompt, and its codes become the vocabulary of exclusion reasons up to the PRISMA diagram. Declared deviations are listed in the Methods draft. The app flags undeclared ones, e.g. criteria edited after screening began.

**Search equations.** Reviewers break each question into concepts with keyword sets, helped by LLM synonym suggestions and MeSH lookups. LitUbik translates each resulting variant into the syntax of 19 databases, with field tags and filters. A protocol's equation can also be pasted as is, and an LLM then drafts its version for the other databases. These drafts stay flagged "not reviewed" until a reviewer validates or edits them, and any query can be written by hand. Hit counts are fetched live wherever an API is reachable. A golden set of must-find references gives each variant a recall score per database, checked against the imported records when there is no API. If no variant is satisfactory, an AI advice button critiques the variants from their counts and recall, without changing them. The chosen equation is frozen with a timestamp and reported in a PRISMA-S search report [2].

**Searching, imports and living review.** Eight open APIs run the frozen equation with pagination: PubMed, OpenAlex [53], Semantic Scholar, Crossref, Europe PMC, arXiv, HAL and DOAJ. Six more (Scopus, IEEE Xplore, Web of Science Starter, CORE, Lens.org and Springer Nature) need a key from the user or their institution, and LitUbik embeds no access of its own. Databases without an API, such as Google Scholar, TRID, EBSCO, the ACM Digital Library and Compendex, follow an assisted-manual path. LitUbik shows the translated equation and export instructions. The reviewer imports the export (RIS, nbib, BibTeX or CSV) with its search date, its displayed hit count and any date or discipline limits, as PRISMA-S asks [2]. Missing abstracts are then fetched from OpenAlex, Semantic Scholar and PubMed. Zotero collections can be imported, and included references flow back to Zotero. A living-review plan re-runs the frozen equation on a schedule, restricted, where the database allows it, to records added since the previous pass. Each pass opens a new round under the same rules, notifies the reviewers in the app and gets its own column of PRISMA counts.

**Deduplication.** Records are deduplicated in three passes: exact DOI or PMID, normalised title, then fuzzy title matching guarded by year and first author. Confident matches merge onto a canonical record that keeps the provenance of each source. Borderline pairs enter a side-by-side arbitration queue. Rulings persist, so deduplication can safely re-run after each import or snowballing round.

**Title/abstract screening.** Each round is configured with its reviewers, its allocation and its conflict rule. Allocation can give each record to everyone, to two reviewers, or to one reviewer with an overlap percentage. It can also follow weighted lots with rotating pairs, or a lead reviewer who screens everything while the others share the corpus. The lot plan can be previewed, edited and exported before the round starts. Conflicts are resolved by consensus or by an arbiter named for the round, by default a project admin. The server withholds each vote from the other reviewers until the record is complete. An option also hides reviewer identities from one another, admins included, until the round ends. The view offers keyboard voting, highlighted protocol terms, notes, tags and marked LLM translations of foreign abstracts. Agreement is computed continuously: Cohen's κ per pair, Fleiss' κ for the team [7,52] and cross-tabulated votes. κ counts a "maybe" as include, but any "maybe" opens a conflict.

AI assistance in screening comes at three levels. A reviewer can request an assessment of one record against the validated criteria (Fig. 3a), and the same assessment can run in batch. A title-only triage proposes clearly off-topic records for exclusion above a confidence threshold. A human reviews these proposals as a checklist before any exclusion takes effect. In an admin mode, the validated proposals resolve records for the whole round. As an option, a control sample of the validated exclusions, from 1% to 50%, returns unmarked to the human queue. The dashboard then shows how many of them the reviewers confirm or contradict. Suggestions are stored apart from votes. A round can also hide them until the reviewer has voted. An earlier look remains possible, but it is recorded and counted in the disclosure.

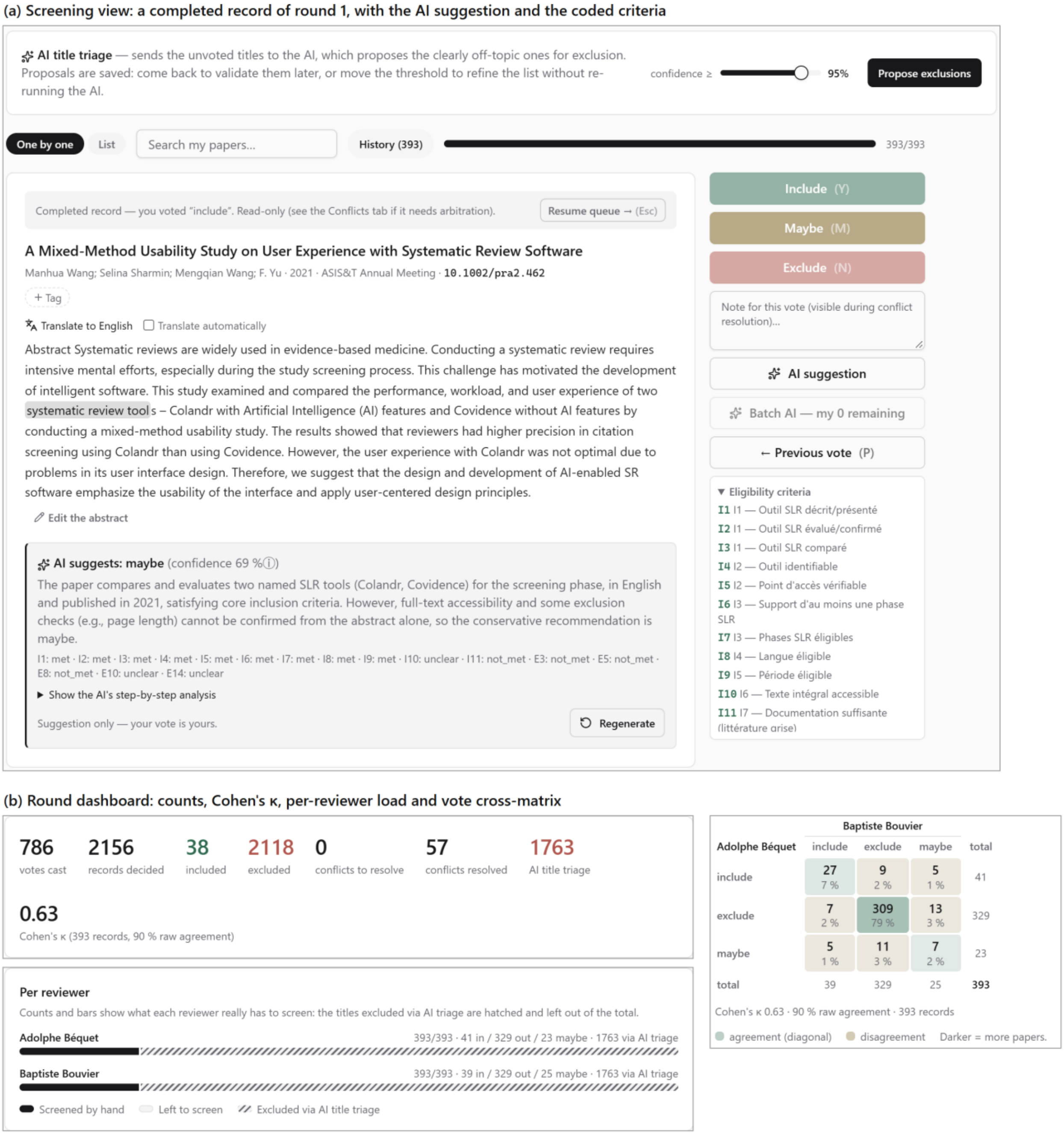

**Figure 3.** Title/abstract screening in the review of Section 3 (round 1). (a) A completed record with the AI suggestion, its criterion-by-criterion assessment and the protocol's coded criteria (labels from the French-language protocol). The bar above sets the confidence threshold of the AI title triage. (b) The round dashboard: counts, Cohen's κ, the load of each reviewer with the share removed by validated AI triage, and the vote cross-matrix.

**Full texts.** Reports retained after title/abstract screening are fetched in bulk. The cascade tries Unpaywall (formerly oaDOI) [54], Europe PMC and open-access publisher pages. It then tries the team's institutional access (EZproxy or CAS) with encrypted credentials. Failures fall back to manual or bulk upload, matched by DOI, file name or title. Reports that cannot be retrieved feed the PRISMA "not retrieved" count. Full-text screening reuses the screening machinery with a built-in PDF viewer. Reviewers can highlight and annotate the PDF, and these annotations are visible to the project members. Excluding a report requires a primary reason from the coded criteria, which gives the PRISMA breakdown by reason without manual counting. Documents over 45 pages are not sent to a model.

**Snowballing.** Backward and forward snowballing [55] harvests Crossref, OpenAlex, Semantic Scholar and OpenCitations [56], plus DOIs mined from stored PDFs. Each iteration deduplicates against the whole corpus and opens a round under the previous rules. It fills missing abstracts, flags saturation and can be undone as a block. Two citation views trace each retained record back to its seed, round by round (Fig. 4). Its records feed the PRISMA "other methods" stream.

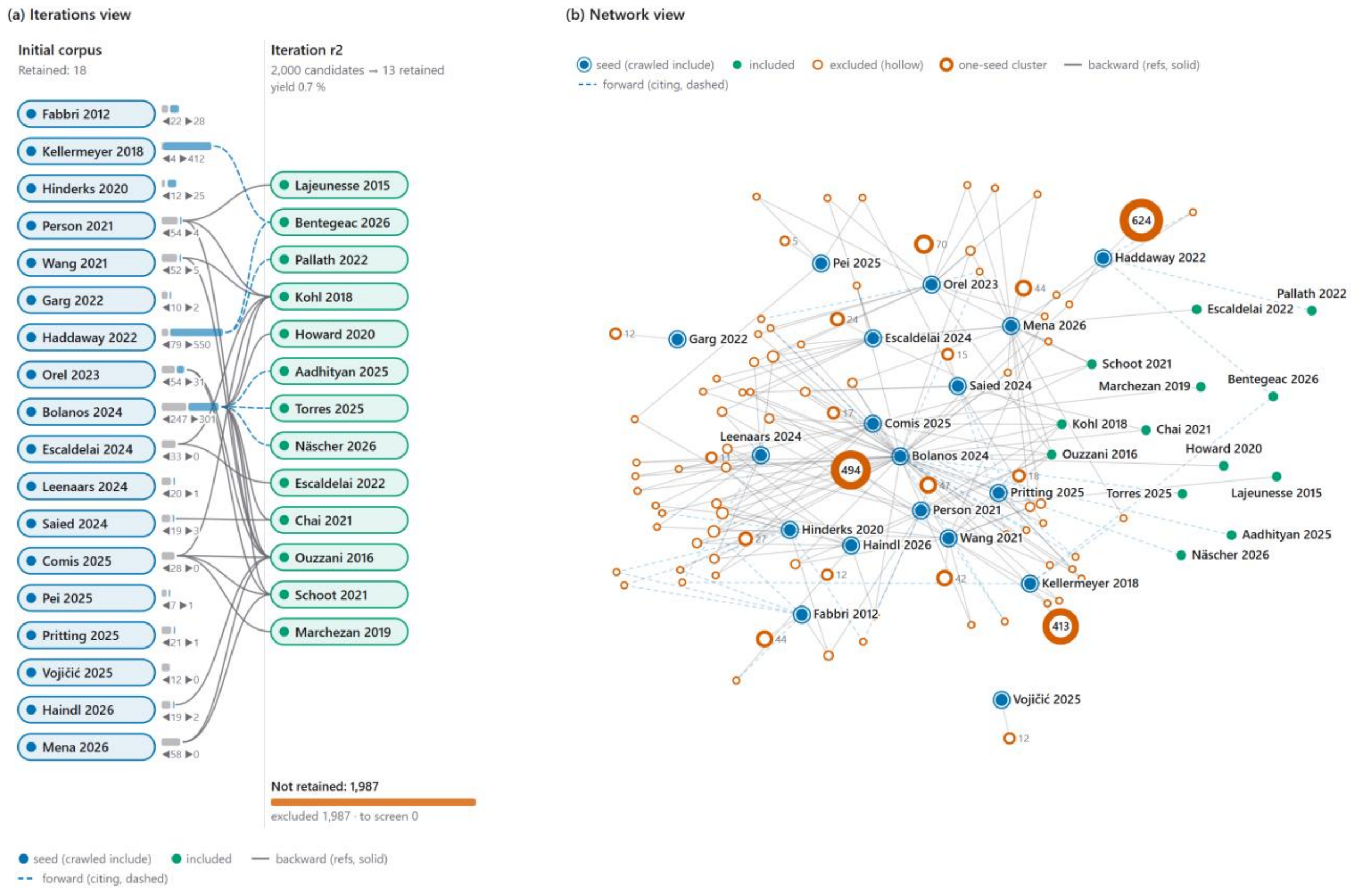

**Figure 4.** Citation network of the snowballing iteration of Section 3, exported from LitUbik: 18 seeds and the 2,000 new records that carry a citation link. (a) Iterations view. Each of the 13 included records is linked to the seeds whose crawl found it. A seed's bar sums up what its crawl returned, and the column gives the round's yield and what was not retained. (b) Network view. Retained records and candidates shared by several seeds are drawn, and the candidates found by one seed only fold into one ring per seed. Solid links are references (backward), dashed links are citing papers (forward).

**Appraisal, extraction and synthesis.** Included studies are appraised with eight JBI critical appraisal checklists [57], with MMAT 2018 [58] or with an instrument the project declares. A report can hold several studies. Appraisal is single or double, with arbitration and a chart per domain. The extraction grid supports typed columns, PICO and SPIDER templates, LLM-proposed columns, import from another platform and double extraction. An LLM can prefill a record's cells from its PDF. Prefilled cells stay "to validate" and never overwrite a human value. A synthesis module records the SWiM items, GRADE certainty per outcome and a summary-of-findings table. A meta-analysis bridge computes per-study effect sizes and exports files for metafor and RevMan. A results page offers charts and a thematic map of the corpus, computed locally.

**PRISMA 2020 and reporting.** The PRISMA page draws the two-stream diagram [1] from live queries, and tabulates each snowballing or update round below it. No count is typed by hand, and the figure exports as SVG or PNG. The reporting stage drafts two texts from the project's records. An LLM writes the Methods draft from the audit trail, from the protocol and frozen equations to κ, retrieval routes and amendments. Missing information is flagged for the author. PRISMA 2020 and SWiM checklists arrive pre-answered. The AI-use disclosure is drafted from the call journal, crossed with what humans did with each output: suggestions followed or overruled, prefilled cells validated, triage proposals confirmed. It adapts the DAISY structure [51], and the author corrects it before generating the statement. The whole project, including every vote with its author and timestamp, exports as a versioned archive that another instance imports to reproduce the review.

**Collaboration across instances.** A team can share one server, run by a lab or online. When collaborators cannot reach a common server, one instance acts as coordinator and the others run as satellites. The coordinator publishes project snapshots. Importing a snapshot creates or updates a satellite, where structural operations are locked on the server. Each satellite exports a contribution package with its votes, appraisals, extraction values, annotations and AI journal. The coordinator merges it idempotently with row-level provenance, quarantines orphans and recomputes conflicts and κ. Tests show the same conflicts, κ and PRISMA as on a single instance. Exchange is automatic through a folder synchronised by Nextcloud, Drive or Dropbox, where a project passphrase authenticates packages and encrypts shared API keys. It can also be manual, with zip files over any channel (Fig. 5).

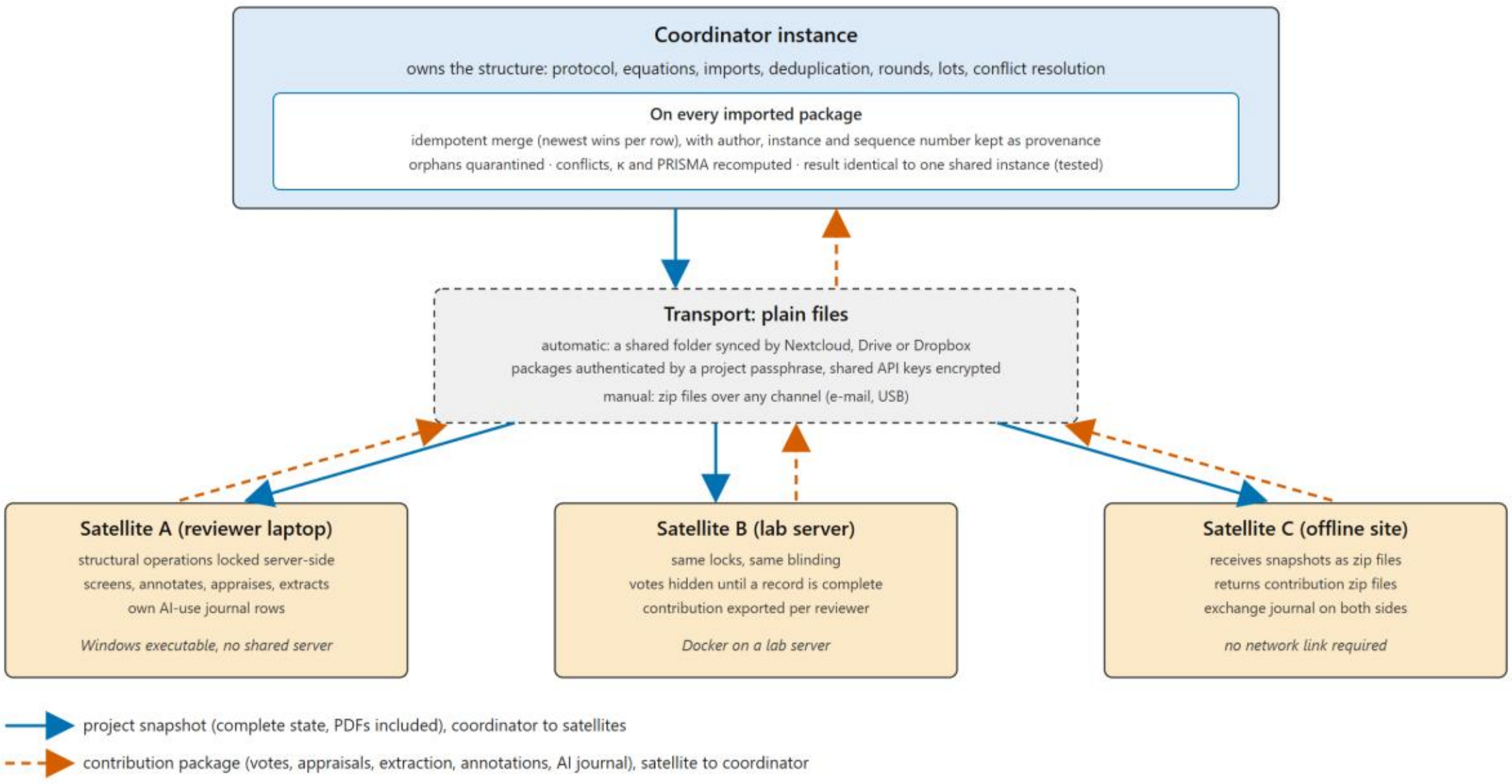

**Figure 5.** Collaboration across disconnected instances. The coordinator publishes project snapshots (solid blue arrows), and satellites return contribution packages (dashed orange arrows), through a shared folder or zip files. Each import is merged with row-level provenance and recomputes conflicts, κ and PRISMA. In shared-server mode, not drawn, everyone works on one instance under the same rules.

## 3. Illustrative example

We used LitUbik to conduct the systematic review of review-support tools that underpins Section 1 and Table 1, with the first and third authors as reviewers. All counts below come from version 5 of the project archive, exported on 21 October 2026. Its public version is provided as supplementary material (S1). Importing it into any LitUbik instance reproduces them.

The two reviewers wrote the protocol offline, then pasted it into LitUbik as text. It has seven research questions on phase coverage, deployment and licensing, AI assistance and human control, methodological safeguards and empirical evidence. The LLM proposed seven inclusion and eleven exclusion criteria, split into 28 coded items (I1–I11, E1–E17). The reviewers revised and validated them on 29 July 2026. One concept ("systematic review tool", eight keyword variants) gave twelve equation variants with live counts. A golden set of three tool papers (Rayyan, StArt, ReviQ) guided the choice, and all three were found in the merged corpus. On 29 July 2026, the frozen equation ran through the PubMed, OpenAlex, Semantic Scholar and Scopus APIs (249, 1,957, 1,492 and 1,000 records). It was run by hand on Web of Science, IEEE Xplore and the ACM Digital Library (397, 25 and 28 records imported). Of these 5,148 records, deduplication removed 2,992 by DOI, normalised title or fuzzy title. The reviewers arbitrated 43 borderline pairs by hand (28 merged, 15 kept apart), leaving 2,156 unique records.

For the title-only AI triage, we chose OpenAI gpt-5 for its balance of quality, cost and speed. It proposed clearly off-topic records for exclusion. The admin reviewed the list and validated 1,763 exclusions (82% of the round), each stored as an audited resolution. Both reviewers screened the other 393 records blind. Raw agreement was 89.8%, with Cohen's κ = 0.63, "substantial" on the Landis and Koch scale [59]. The 57 conflicts were resolved by discussion between the reviewers (46 excluded, 11 included), and 38 records went to full text. Full-text screening included 18 of these 38 reports (κ = 0.79, substantial). Each exclusion carried a primary coded reason. The most frequent were an inaccessible full text (8 reports), an ineligible publication type such as extended abstracts or posters (4), a duplicate version (2) and a purely algorithmic contribution (2). Two amendments were declared. First, many papers evaluated tools already presented elsewhere in the corpus, so inclusion was restricted to papers that present a tool. The second concerns full texts that could not be obtained (below). Both appear in the register and in the Methods draft.

Over the whole review, 62 reports were sought. LitUbik retrieved 38 automatically: 20 through Unpaywall, 12 through the institutional proxy and 6 from publisher or Europe PMC pages. The reviewers uploaded 20 by hand. Four full texts could not be obtained, because we had no access to their publisher and their authors could not be reached. Six retrieved files were wrong or corrupt documents. All ten were excluded for inaccessible full text, the first four through a declared amendment.

The 18 included reports seeded one snowballing iteration on 24 September 2026. The four citation sources returned 788 backward and 1,434 forward citations, giving 2,104 candidate records. Of these, 103 duplicated the corpus, and the 2,001 new records opened round 2 under the same rules. AI triage with gpt-6-astra, validated by the admin, excluded 1,185 of them. The two reviewers double-screened the other 816 (agreement 93.1%, κ = 0.49, moderate). They resolved 79 conflicts (63 excluded, 16 included) and sent 24 reports to full text. Thirteen were included (κ = 0.83, almost perfect). The review thus comprises 31 studies, 13 of them found only by snowballing. A second iteration, which would test saturation, is planned. Figure 6 shows the PRISMA 2020 diagram exported from the project. An LLM prefilled a 36-column extraction grid (S2), which the reviewers checked and which supports Table 1. The archive keeps its cells awaiting validation, to show the suggestion interface.

The AI-use journal records 302 calls to three OpenAI models across nine tasks, from protocol extraction to extraction-column design (Table A1). Several models were tried on the same records to check consistency. The calls used 1.45 million input and 0.50 million output tokens, for an estimated USD 29.60 at current list prices. Crossed with the human decisions, the journal gives the material of the disclosure. Of 204 title/abstract suggestions, all 67 "exclude" suggestions matched the final decision, and 20 of the 137 "maybe" suggestions ended included. Of 45 full-text suggestions, 38 (84%) matched. In the 7 others, the reviewers excluded reports that the AI rated "include" or "maybe". A human reviewed each of the 2,948 triage exclusions before it took effect. These matches do not prove independent agreement. Here, suggestions were visible before the vote, so a reviewer could follow them. Rounds can now hide suggestions until the vote, which makes this distinction measurable.

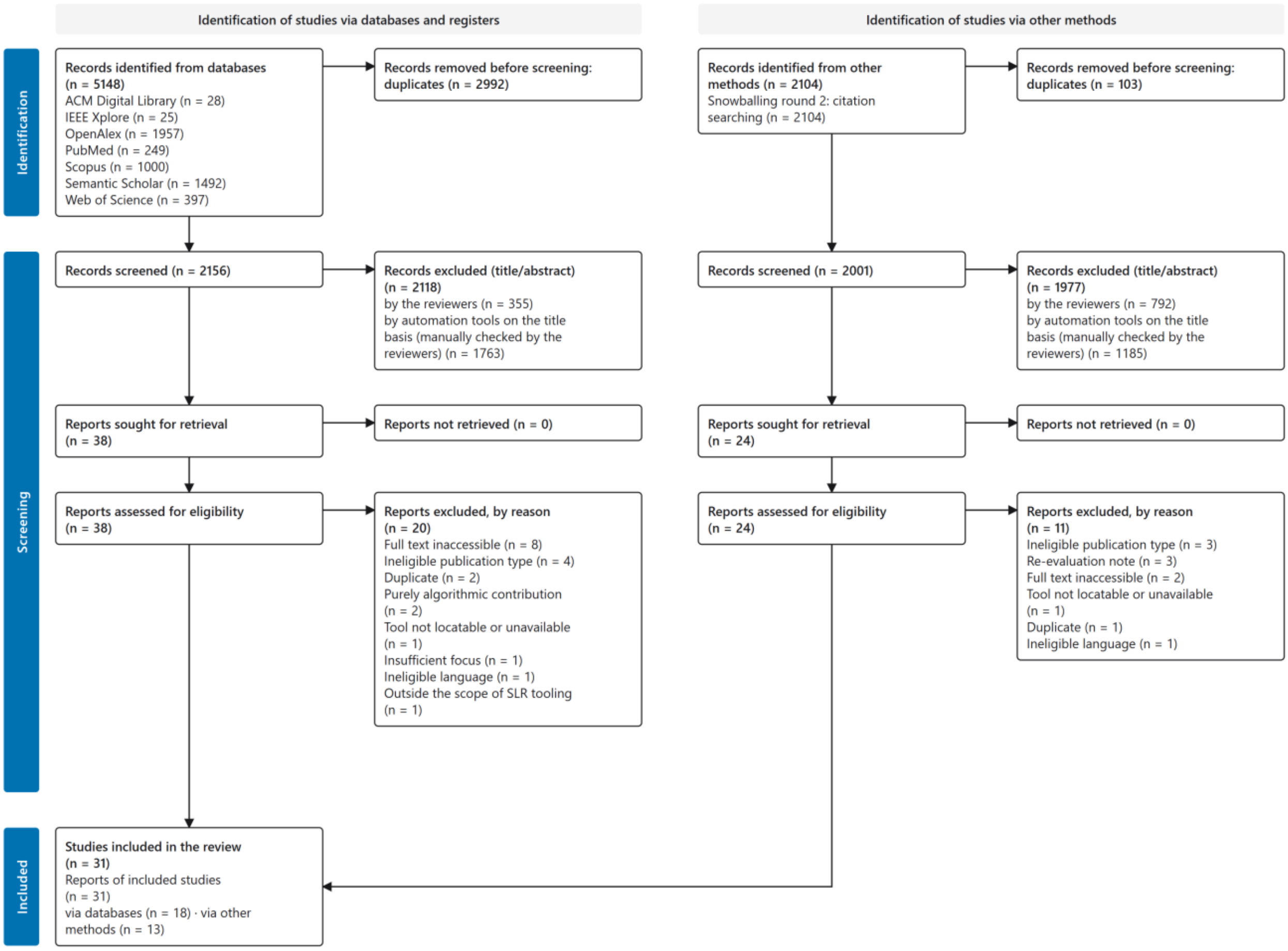

**Figure 6.** PRISMA 2020 flow diagram of the review of Section 3, exported by LitUbik from version 5 of the project archive (21 October 2026). It shows the two identification streams and the exclusions by coded reason. Reason labels are translated from the French-language protocol.

## 4. Impact

LitUbik changes what a review team can do in four ways.

**AI assistance under human control.** LitUbik takes a clear position in the debate on LLMs in evidence synthesis [14–17]. It offers AI help at every stage, but each output is a proposal that a named reviewer accepts, edits or rejects. Rounds can hide suggestions until the reviewer has voted. This answers the anchoring risk that LitFlow's test users raised [26]. The journal counts each billed call, so the disclosure statement is written from a record rather than from memory. This meets the expectations of the evidence-synthesis organisations [19] and avoids relying on self-reports [20]. To our knowledge, no other review tool logs its own AI use for this purpose. The local-provider option keeps embargoed material on the team's machines. The journal also travels in contribution packages, so a coordinator's statement covers AI used on satellites.

**One consistent record.** ReviQ showed that deriving PRISMA counts from the decision log keeps them consistent with the decisions [12]. LitUbik extends this to the stages before screening, because it also holds the equations, database runs, import files and deduplication log. A changed decision, such as an unmerged pair, a re-resolved conflict or a new iteration, updates every statistic, figure and draft. The golden-set mechanism turns search design into a documented calibration. In our review, the closest tools offered only search-string metrics or precision and recall computed after screening [12,25].

**Collaboration without shared infrastructure.** On one server, a small team gets server-enforced blinding, live Cohen's and Fleiss' κ, weighted lots and auditable conflict resolution, on infrastructure it controls. When collaborators cannot share a server, the coordinator–satellite exchange transports the complete state and each reviewer's contributions with provenance. Merging recomputes conflicts, κ and PRISMA under the project's rules, so the exchange cannot bypass the methodology. It also removes the scaling limit reported for ReviQ's JSON exchange [12].

**A dataset on AI assistance.** Each completed review pairs human decisions and model assessments over the same records, under the same protocol, with confidence and rationale. These are the data needed to evaluate LLM screening in real conditions [18]. With the journal and the declared statement in one system, they also allow comparing what was declared with what was done, as DAISY's authors propose [51]. Section 3 gives a first such dataset.

The authors' laboratory maintains LitUbik. Releases are built by continuous integration, issues are tracked on GitHub, and in-app feedback reaches the instance administrator. Four beta testers outside the development team have used it and reported on its usability.

**Limitations.** LitUbik has no local active-learning prioritisation and no recall estimation. ASReview offers the former [23], and SWIFT-Active Screener, a web service, adds recall estimation to decide when to stop [30]. We gave priority to LLM assistance, which is more recent. An LLM works from the validated criteria and returns a rationale per criterion that the reviewer can check. Closed databases without an API key still need the assisted-manual path. The κ of 0.49 in round 2 of Section 3 shows that a broad snowballing corpus lowers agreement, which the tool measures but cannot fix. SQLite suits small teams, and a PostgreSQL backend is future work. The packaged Windows executable omits the browser-based and institutional retrieval routes, which need the source or Docker installation.

## 5. Conclusions

LitUbik is open-source software in which humans conduct a systematic review and AI assists them at every stage, under systematic human validation and with full provenance. To our knowledge, it is the only open-source application that covers the full review lifecycle, from the registered protocol to the Methods and AI-disclosure drafts, in one self-hosted system. Collaboration works with the same guarantees on a shared server or across disconnected instances. A systematic review of 31 tool papers, conducted in LitUbik, positions it against existing tools and shows it at work on a real review. Future work includes recall estimation for prioritised screening, a second real-world review with a larger team and a usability study with the System Usability Scale. We invite the community to use, audit and extend LitUbik. Its name blends literature with Ubik, Philip K. Dick's novel [60] in which reality drifts unless actively held together, a fair description of a living review corpus.

## Declaration of competing interest

The authors declare that they have no known competing financial interests or personal relationships that could have appeared to influence the work reported in this paper.

## Declaration of generative AI and AI-assisted technologies in the manuscript preparation process

During the preparation of this work, the authors used OpenAI GPT models, through LitUbik, to triage titles, suggest screening decisions and draft search concepts in the review reported in Section 3, under the human validation described above. They used Anthropic Claude, through Claude Code, to help write the software, prefill the data-extraction grid from the included papers, draft the manuscript text and diagrams, and check citations against their sources. After using these tools, the authors reviewed and edited the content as needed and take full responsibility for the content of the publication.

Table A1 details this use with the DAISY structure [51]: level of support per activity, tools and versions, material shared with the provider, and human validation. Part A is read from LitUbik's AI-use journal for the review of Section 3 (archive of 21 October 2026), except the extraction grid, prefilled outside LitUbik. Parts B and C are declared by the authors.

**Table A1.** AI use by activity. Levels: none, low, medium, high.

| Activity | Level | Tool and version | Material shared with the provider | Human validation |
|---|---|---|---|---|
| **A. Conducting the review of Section 3 (journaled by LitUbik)** | | | | |
| Protocol structuring and eligibility criteria | low (1 call) | OpenAI gpt-5 via API | the pasted protocol text | 28 coded criteria proposed by the model, revised and validated by the reviewers on 29 July 2026 |
| Search strategy design (concept proposal, synonyms, equation advice) | low (4 calls) | OpenAI gpt-5 | research question, protocol, concept list | proposals added to an editable keyword list. The twelve variants were written and frozen by the authors |
| Study selection: bulk title triage | high (2,948 of 4,157 title/abstract records excluded by validated triage) | OpenAI gpt-5 (round 1), gpt-6-astra (round 2). 43 calls | titles only, with the validated protocol | each proposal reviewed in a checklist by the project admin before taking effect, and stored as an audited resolution. No proposal applied without review. No control sample (option added later) |
| Study selection: title/abstract suggestions | low (204 of 4,157 records, 203 calls) | OpenAI gpt-5, gpt-5.6-terra, gpt-6-astra | title, abstract, protocol criteria | stored apart from votes, visible before the vote in this review. All 67 "exclude" suggestions matched the final decision. 20 of 137 "maybe" suggestions ended included |
| Study selection: full-text suggestions | medium (45 of 62 reports, 45 calls) | OpenAI gpt-5, gpt-6-astra | text of the PDF (≤ 45 pages) and protocol | 38 of 45 matched the final decision. In the 7 others, the reviewers excluded reports that the model rated include or maybe |
| Reading aid: abstract translation | low (4 calls) | OpenAI gpt-6-astra | title and abstract | shown next to the original, marked "AI translation" |
| Data extraction: column design | low (2 calls) | OpenAI gpt-6-astra | protocol and research questions | proposals not kept. The final grid follows the rows of Table 1. No cell prefilled by AI within LitUbik |
| Data extraction: filling the grid (31 reports, 36 columns), outside LitUbik and not journaled | high | Anthropic Claude Opus 5.5 through Claude Code, October 2026 | the PDFs of the 31 included reports | each value proposed with its page and quote, then checked by both reviewers. Imported as AI proposals and left awaiting validation in the archive, so that readers can see the suggestion interface |
| Critical appraisal, synthesis, reporting drafts | none | | | |
| Totals | 302 calls, 1.45 M input and 0.50 M output tokens, about USD 29.60 at current list prices (the archive's journal shows USD 6.78, because two of the models were missing from LitUbik's price table and were priced at gpt-5 rates), 29 July to 12 October 2026. Several | | | |

| Activity | Level | Tool and version | Material shared with the provider | Human validation |
|---|---|---|---|---|
| | models were tried on the same records to check consistency. | | | |
| **B. Developing the software** | | | | |
| Code generation, refactoring, tests and documentation | high | Anthropic Claude through Claude Code: Opus 4.8, Opus 5, Opus 5.5, Fable 5 and Fable 5.1, July to October 2026 (co-author lines of the commits of the development repository) | source code of the repository and the first author's specifications | each change reviewed by the first author, covered by the backend test suite and the axe-core audit, and committed under the author's name with the assistant credited as co-author |
| **C. Preparing this manuscript** | | | | |
| Drafting of the text from the authors' plan, the feature matrix and the project export | high | Anthropic Claude through Claude Code: Fable 5.1 (drafts v0.1 to v0.3), Opus 5.5 (v0.4 to the submitted version), October 2026 | writing plan, feature matrix, project archive, repository | two review rounds by the authors (125 comments on v0.3). The authors rewrote the text as needed, checked every number against the project export and take full responsibility for the content |
| Checking citations against the source PDFs | medium | Anthropic Claude Opus 5.5 | the cited PDFs | each verdict given with the page and exact quote of the source. Every resulting correction approved by the authors |
| Figures 1, 2 and 5 (diagrams) | high | Anthropic Claude (as above) | the authors' specification | checked and corrected by the authors |
| Figures 3 and 4 | none (screenshots of the application) | | | |
| Figure 6 | low | Anthropic Claude Opus 5.5 | the PRISMA export of the application | reason labels translated into English and counts set to archive version 5 by a script. Every count checked against the PRISMA computation of that archive |

**Safeguards.** API keys are stored encrypted on the LitUbik server. Calls went to the OpenAI API under its API data-usage terms (no training on API inputs). PDFs over 45 pages are never sent. The prompts are readable and editable in the project settings. In this review, every call made within LitUbik went through the logged chain, which counted each completed call per task, model, user and day. That version did not yet count calls refused by the model or attempts that failed before a provider failover. LitUbik 1.0.0 counts them. The extraction grid was prefilled outside LitUbik, so it is declared separately in Table A1. AI tools are not authors, and the authors take full responsibility for the review, the software and the manuscript.

## Data availability

The source code is available at https://github.com/abequet/litubik_public under the AGPL-3.0-or-later license (version 1.0.0), and archived on Zenodo (https://doi.org/10.5281/zenodo.23652584 for this version, https://doi.org/10.5281/zenodo.23652583 for all versions). The public replication archive of the review of Section 3 is provided as supplementary material on Zenodo (https://doi.org/10.5281/zenodo.23662137). LitUbik exports it without the full texts, abstracts and raw database exports, which belong to publishers and database vendors, and without the shared API keys or the reviewers' work sessions. It keeps the metadata, decisions, reasons, extraction and AI-use journal, so importing it reproduces every count of Section 3. Whoever replicates the review retrieves the full texts with their own access. The extraction grid that supports Table 1 is provided as supplementary material (S2), with the page and quote behind each value. S1 and S2 are released under CC BY 4.0, except the bibliographic metadata, which remains subject to the terms of the databases searched. The two reviewers, both authors, agreed to be named in S1. A public demonstration instance runs at https://demo.litubik.org: one click opens an anonymous guest session, without an account, on an example review already set up. Guest sessions are deleted daily, retrieve full texts from open-access sources only, and run AI assistance within a capped budget.

## CRediT authorship contribution statement

**Adolphe J. Béquet:** Conceptualization, Methodology, Software, Validation, Investigation, Data curation, Visualization, Writing – original draft, Writing – review & editing, Project administration. **Sébastien Laurent:** Software (user-interface design), Writing – review & editing. **Baptiste Bouvier:** Investigation (second reviewer in Section 3), Validation, Writing – review & editing.

## Funding

LitUbik was developed to support the systematic review of the RESCOUSSE project (WP1), funded by the French Road Safety Delegation (DSR), grant no. 2201600314.

## Acknowledgements

We thank our beta testers, Morgane Lacroix, Noé Marie, Maud Ranchet and Caroline Pigeon, for their valuable feedback and usability tests.